\documentstyle[12pt,epsfig]{article}
\newcommand{\be}[1]{
\begin{eqnarray}\label{#1}}
\newcommand{\ee}{\end{eqnarray}}
\newcommand{\ci}[1]{\cite{#1}}
\newcommand{\re}[1]{(\ref{#1})}

\newcommand\pbar{\bar{\psi}}
\newcommand\p{\psi }

\newcommand{\spin}[1]{ \langle\mskip-3mu \langle{#1}
\rangle\mskip-3mu\rangle}

\newcommand{\eps}{\epsilon}

\newcommand{\inclfig}[2]{\mbox{\epsfxsize=#1cm \epsfbox{#2.ps}}}

\begin{document}
\renewcommand{\thefootnote}{\fnsymbol{footnote}}
\begin{flushright}
\begin{tabular}{l}
 RUB-TPII-03-07\\
\end{tabular}
\end{flushright}
\begin{center}
{\bf\Large NLO corrections to the twist-3
   amplitude in DVCS on a nucleon in the Wandzura-Wilczek
  approximation: quark case }

\vspace{0.5cm} N. Kivel$^{a}$\footnote{on leave of absence from
St.Petersburg
Nuclear Physics Institute, 188350, Gatchina, Russia },
L. Mankiewicz$^b$,

\begin{center}
{\em $^a$ Institut f\"ur Theoretische Physik II,
Ruhr-Universit\"at Bochum,\\ D-44780 Bochum, Germany}
\\
{\em $^b$ Center for Theoretical Physics, Polish Academy of Science,
Al. Lotnik\'ow 32/46,  02-668 Warsaw, Poland}

\end{center}

\end{center}

\begin{abstract}
We computed the NLO corrections to twist-3, $L \rightarrow T$,
flavor non-singlet amplitude in DVCS on a nucleon in the Wandzura-Wilczek
approximation. Explicit calculation shows that factorization holds
for NLO contribution to this amplitude, although the structure of
the factorized amplitude at the NLO is more complicated than in
the leading-order formula. Next-to-leading order coefficient functions
for matrix elements of twist-3 vector and axial-vector quark string
operators and their LO evolution equations are presented.
\end{abstract}

\newpage
\section*{\normalsize \bf Introduction}

Deeply virtual Compton scattering (DVCS) \cite{DVCS1,DVCS2} on a nucleon,
$\gamma^* N \rightarrow \gamma N'$ , is perhaps the
cleanest hard reaction sensitive to the skewed parton distributions
(SPD). According to celebrated factorization theorems
\cite{Rad97,Ji98,Col99} the leading term
in the $1/Q^2$ expansion of
the DVCS
amplitude, where $Q^2$ is
large virtuality of the hard photon, can be
expressed in terms of twist-2 skewed parton distributions.
Twist-2 SPD's are related to matrix elements of non-local twist-2 quark
and gluon string operators. These matrix elements contain complete
information about nucleon structure seen by
a highly-virtual electromagnetic probe. For that reason in recent
years DVCS has been the subject of extensive
theoretical investigations. First experimental data have also became recently
available (see e.g. \cite{exp1,exp2,HermesSSA, amarian}) and much more data
are expected
from JLAB, DESY, and CERN in the future.

The amplitude of $\gamma^* N \rightarrow \gamma N'$ scattering
receives contribution from both DVCS and Bethe-Heitler processes.
Extraction of DVCS from the experimental cross-section is
therefore not straightforward. The helicity analysis reveals that
the interference term contains contributions of different photon
helicity amplitudes weighted with sines and cosines of the angle
between leptonic and hadronic scattering planes in the
center-of-mass of the scattered photon and the nucleon
\ci{PVG,BMK}. At present, the leading-twist DVCS amplitude has
been studied to the NLO accuracy \cite{JiOs,BMnlo,MPS}. The
analysis has revealed that the photon helicity is conserved by the
tree-level contribution. Photon helicity-flip amplitude appears at
the NLO level. The $L \rightarrow T$ amplitude, corresponding to
scattering of longitudinal virtual photon off nucleon, appears as
twist-3, $1/Q$ correction. As the typical experimentally
accessible values of $Q^2$ are by no means large, such leading
power corrections may have significant effects on some of DVCS
observables. In addition, twist-3 corrections typically scale as
$\sqrt{-t}/Q$, with $t$ denoting the square of the momentum
transfer, so the size of twist-3 corrections increases with $t$.
As it follows, taking into account these corrections is mandatory
for understanding continuation of the twist-2 part of the DVCS
amplitude to $t=0$.

The LO contribution to the DVCS amplitude on the nucleon to the
twist-3 accuracy has been calculated in \cite{Penttinen,BM}.
Twist-3 corrections to the DVCS amplitude arise in a twofold way.
First, as usual for a hard exclusive process, the hard amplitude
with an additional parton, the gluon, taking part in the hard
collision is suppressed by one power of $1/Q$. Such a genuine
twist-3 contribution involves matrix element of three-parton
operator in a nucleon state. Second, there is a so-called
Wandzura-Wilczek (WW) contribution which arises from a hard
configuration with minimal number of hard partons involved in hard
collision. Formally, the WW contribution arises because  operators
with external derivatives w.r.t. total translation in a transverse
direction give nonzero contribution in the DVCS kinematics. As the
current phenomenology of power corrections is consistent with an
assumption that matrix elements of three-parton operators in a
nucleon are small \ci{Kipt}, one can conjecture that the WW
contribution can provide a rather accurate numerical description
of  twist-3 corrections.

In this paper we have computed the NLO contribution to the WW
twist-3, $L \rightarrow T$ flavor non-singlet DVCS amplitude.
Besides obvious phenomenological applications, there is a broader,
theoretical interest in such a calculation. A factorization
theorem for twist-3 DVCS amplitude has neither been considered nor
proven in the literature. As a consequence, although the direct
calculation showed that the LO $L \rightarrow T$ amplitude
factorizes, there is no guarantee that factorization prevails in
higher orders as well. At the same time there have been no direct
calculations supporting or disproving factorization of twits-3
contribution to DVCS beyond the leading order. The aim of this
paper has been to provide an explicit example of NLO correction to
twist-3 WW amplitude. Although computation of one-loop, flavor
non-singlet amplitude may seem trivial, based on experience with
NLO corrections to twist-2 amplitudes, calculation of NLO twist-3
contribution has turned out to be technically quite involved,
warranting a more detailed discussion. Our calculation
demonstrates that the amplitude does factorize at the NLO. It is
tempting to interpret it as a hint that the factorization holds
for this particular twist-3 DVCS helicity amplitude to all orders.

Our paper is organized in the following way: in the first section we
introduce our definitions and notations. The next two sections
are devoted to discussion of intricacies of the calculation: in the second
section we derive convenient parametrization of matrix elements of
vector and
axial-vector quark string operators up to twist-3 accuracy in the WW
approximation and in the third one we demonstrate technicalities of
our approach by calculating LO twist-3 term in WW approximation. The
fourth section contains discussion of our main result - the NLO
correction to the WW $L
\rightarrow T$ amplitude. Finally, we conclude.

\section*{\normalsize \bf DVCS amplitude on a nucleon }

Let $p, p'$ and $q,q'$ denote momenta of the initial and final
nucleons and photons, respectively. The amplitude of the virtual
Compton scattering process \be{proc} \gamma^*(q)+N(p)\to
\gamma(q') +N(p') \, , \ee is defined in terms of the nucleon
matrix element of the $T$-product of two electromagnetic
currents~: \be{T:def} T^{\mu\nu}=i\int d^4x\ e^{-i
(q+q')x/2}\langle p'|T\left[J_{\rm e.m.}^\mu (x/2) J_{\rm
e.m.}^\nu(-x/2)\right]|p\rangle\, , \ee where Lorentz indices
$\mu$ and $\nu$ correspond to the virtual, respectively the real
photon.

We shall consider the Bjorken limit, where $-q^2=Q^2\to\infty$,
$2(p\cdot q)\to \infty$, with $x_B = Q^2 / 2(p\cdot q)$ constant,
and $t \equiv (p-p')^2\ll Q^2$.
We introduce two light-like vectors $n,\, n^*$ such that
\be{nnstar} n\cdot n=0, \, n^*\cdot n^*=0, n\cdot n^*=1. \ee We
shall work in a reference frame where the average nucleon momenta
$P =\frac12(p+p')$ and the virtual photon momentum  $q$ are
collinear along z-axis and have opposite directions. Such a choice
of the frame results in the following decomposition of the momenta
\ci{PVG}:
\be{kinem}
P&=& n^* + \frac{{\bar m}^2}{2} n \, \nonumber
\\[4mm]
q&=& -2\xi^\prime n^* +\frac{Q^2}{4\xi^\prime}n \, \nonumber
\\[4mm]
\Delta&=& p'-p = - 2 \xi n^* + {\bar m}^2 \xi n + \Delta_\perp
\ee
with ${\bar m}^2 = m^2 - t/4$, $t = \Delta^2$ being the squared momentum
transfer, and
\be{xi}
2 \xi = 2 \xi^\prime \frac{Q^2-t}{Q^2 + 4 \xi^{\prime 2} {\bar m}^2} \, .
\ee
Finally, $\xi^\prime$ is given by
\begin{equation}
\label{xiprime}
\xi^\prime
= \frac{2}{\frac{2-x_B}{x_B} + t/Q^2 + \sqrt{(\frac{2-x_B}{x_B} + t/Q^2)^2
    + 16 \frac{ {\bar m}^2}{Q^2}}} = \frac{x_B}{2 - x_B} + O(1/Q^2) \, ,
\end{equation}
with $x_B = \frac{Q^2}{2 p \cdot q}$.

We define the transverse metric and
antisymmetric transverse epsilon tensors
\footnote{The Levi-Civita
tensor $\epsilon_{\mu \nu \alpha\beta}$ is defined as the totally
antisymmetric tensor with $\epsilon_{0123}=1$ }:
\be{gt} g^{\mu
\nu}_\perp = g^{\mu \nu}- n^\mu n^{*\, \nu}-n^\nu n^{*\, \mu},
\quad \epsilon^\perp_{\mu \nu}= \epsilon_{\mu \nu
\alpha\beta}n^\alpha n^{*\,\beta}\, .
\ee
In the following, we shall use the shorthand notation for
\be{defdot}
 a^+\equiv a_\mu n^\mu, \quad  a^-\equiv a_\mu n^{*\, \mu}\, ,
\ee
where $a_\mu$ is an arbitrary Lorentz vector.

In this paper we focus our attention on the L$\rightarrow$T DVCS
amplitude where the virtual photon has longitudinal polarization. It
appears first at the twist-3 level.\footnote{We adopt here kinematical
definition of twist i.e., terms suppressed by $1/Q$ are of twist-3.}
In order to calculate the leading contribution to this amplitude it is
sufficient to expand (\ref{kinem}) and (\ref{xi}) to twist-3 accuracy:
\be{lce}
P&=&\frac12(p+p')=n^*, \quad \Delta = p'-p =-2\xi
P+\Delta_\perp, \, \nonumber
\\[4mm]
q&=&-2\xi
P+\frac{Q^2}{4\xi}n, \quad
q'=q-\Delta=\frac{Q^2}{4\xi}n-\Delta_\perp\, .
\ee

The LO L$\rightarrow$T amplitude has the form \cite{Penttinen,BM}~:
\be{T} T^{ \mu \nu}_{0+}= -\frac{(q+4\xi P)^\mu}{(Pq)}
\left[g^{\nu \alpha}_\perp+\frac{P^\nu\Delta_\perp^\alpha}{(Pq)}
\right] \frac12\int_{-1}^1 dx\,\left\{ F_\alpha(x,\xi)C^+(x,\xi)-
i\epsilon^\perp_{\alpha \rho}\widetilde F^\rho (x,\xi)
C^-(x,\xi)\right\}  \, . \ee
Note that the combination $(q+4\xi P)^\mu$ is proportional to the
longitudinal polarization vector of the initial photon. In order
to have exact gauge invariance of the amplitude we keep in \re{T}
terms of order $\Delta_\perp^2/Q^{2}$ applying prescription
$g^{\nu \alpha}_\perp\rightarrow g^{\nu
\alpha}_\perp+\frac{P^\nu\Delta_\perp^\alpha}{(Pq)}$ for the
twist-3 terms in the amplitude. The skewed parton distribution
$F_\mu(x,\xi)$ and $\widetilde F_\mu(x,\xi)$ can be related to the
twist-2 SPD's $H,E,\widetilde H,\widetilde E$ with help of WW
relations. We shall give explicit expressions for them in the next
section. Term proportional to $\frac{P^\mu\Delta_\perp^\nu}{(Pq)}$
is required to ensure electromagnetic gauge-invariance: the
amplitude \re{T} is electromagnetically gauge invariant:
\be{Ginv1}
q_\mu T^{\mu\nu}_{0+}= T^{\mu\nu}_{0+} q'_\nu = 0\,
\ee
formally to the accuracy $1/Q^2$. In the LO in QCD coupling
coefficient functions are given by
simple expression which have poles at $x=\pm \xi$:
\be{CF}
\nonumber
C^\pm(x,\xi)=\frac{1}{x-\xi+i\varepsilon}\pm
\frac{1}{x+\xi-i\varepsilon}\, .
\ee
It was demonstrated in \ci{KPST} that despite presence of these
poles the convolution integral in (\ref{T}) is well-defined and
therefore factorization is not violated at the LO.

One final comment is in order here. Note that
using symmetry of the GPD's, see \re{asym} below,
the LO amplitude can be written in a shorter form as
\be{Tshort} T^{ \mu \nu}_{0+}&=& \frac{(q+4\xi P)^\mu}{2(Pq)}
\left[g^{\nu \alpha}_\perp+\frac{P^\nu\Delta_\perp^\alpha}{(Pq)}
\right] \int_{-1}^1 dx\, \frac{-2}{x+\xi} \biggl(
F_{\perp\alpha}(x,\xi)+ i\epsilon^\perp_{\alpha \rho}\widetilde
F^\rho_\perp (x,\xi)\biggr) \, . \ee
In the following we shall, for notational simplicity, always rewrite
lengthy amplitudes in a
similar way, making
use of symmetries of corresponding
GPD's. One can easily make use of these symmetries to transform
expressions into the original form.

\section*{\normalsize \bf Light-cone expansions of the  matrix element }

Calculation of NLO corrections to the L$\rightarrow$T amplitude
requires knowledge of the nucleon matrix element of vector and
axial-vector non-local quark string operators
\be{Voper}
\langle p'|
\pbar(x)\gamma^\sigma\p (y)
 |p\rangle
\ee
and
\be{Aoper}
\langle p'|
\pbar(x)\gamma^\sigma\gamma_5\p (y)
|p\rangle
\ee
to the twist-3 accuracy but for arbitrary $x$ and $y$. We shall
restrict ourselves to the Wandzura-Wilczek approximation i.e., we
neglect quark-gluon-quark operators arising in the twist
expansion. As we shall show below, the result is given in terms of
distribution functions which parametrize matrix elements of string
operators restricted to the light-cone:
\be{Fdef}
F_\mu(x,\xi)=\int^\infty_{-\infty}
\frac{d\lambda}{2\pi}e^{-ix\lambda} \langle p'|
\pbar(\frac12\lambda n)\gamma_\mu\p (-\frac12\lambda n)
 |p\rangle \, ,
\ee
\be{Ftlddef}
\widetilde
F_\mu(x,\xi)=
\int^\infty_{-\infty}\frac{d\lambda}{2\pi}e^{-ix\lambda} \langle p'|
\pbar(\frac12\lambda n)\gamma_\mu\gamma_5\p (-\frac12\lambda n)
 |p\rangle \, .
\ee To simplify notation, we have omitted in the above formulae
the path-ordered exponential connecting the quark fields.

As discussed in \cite{BM,KP}, to the twist-3 accuracy the above
matrix elements can be parametrized as:
\be{def:FWW}
F^{WW}_\mu(x,\xi)&=& P_\mu F(x,\xi)+ F_{\perp\mu}(x,\xi)\, .
\ee
\be{def:tFWW}
\widetilde F^{WW}_\mu(x,\xi)&=& P_\mu
\widetilde F(x,\xi)+ \widetilde F_{\perp\mu}(x,\xi)\, .
\ee
Here we have explicitly separated longitudinal, twist-2:
\be{def:F} F(x,\xi)= n^\rho F^{WW}_\rho(x,\xi) =
\spin{\gamma^+}(H+E)(x,\xi)-\spin{\frac1 m}E(x,\xi)\, , \ee
\be{def:tF} \widetilde F(x,\xi)= n^\rho\widetilde
F^{WW}_\rho(x,\xi)=\spin{\gamma^+\gamma_5}\widetilde H(x,\xi)-
\spin{\frac{\gamma_5}{m}}\xi\widetilde E(x,\xi)\, , \ee
and transverse, twist-3 components:
\be{def:Fperp} F_{\perp\mu}(x,\xi)=
-\frac{\Delta_{\perp\mu}}{2\xi}F(x,\xi)+
\int_{-1}^{1}du\left( \
G_\mu(u,\xi)W_{+}(x,u,\xi)+i\epsilon_{\perp \mu \alpha}\widetilde
G^\alpha (u,\xi)W_{-}(x,u,\xi)\right)\, , \ee
\be{def:tFperp} \widetilde F_{\perp\mu}(x,\xi)=
-\frac{\Delta_{\perp\mu}}{2\xi}\widetilde F(x,\xi)
+\int_{-1}^{1}du\left(  \widetilde G_\mu(u,\xi)W_{+}(x,u,\xi)+i
\epsilon_{\perp \mu \alpha} G^\alpha (u,\xi)W_{-}(x,u,\xi)
\right)\, . \ee
A shorthand notation
$\spin{\ldots}$ denotes $\bar U(p')\ldots U(p)$ and $m$ is the nucleon mass.

Functions $G^\mu$ and $\widetilde G^\mu$ and the Wandzura-Wilczek
kernels $W_{\pm}(x,u,\xi)$
have been introduced in Refs.~\cite{BM,KP}. They are defined as~:
\be{G}
G^\mu(u,\xi)&=& \spin{\gamma^\mu_\perp}(H+E)(u,\xi)-
\frac{\Delta_\perp^\mu}{2\xi}\biggl[u\frac{\partial}{\partial u}+
\xi\frac{\partial}{\partial \xi} \biggl] F(u,\xi) \, , \ee
\be{tG} \widetilde G^\mu (u,\xi)&
=&\spin{\gamma^\mu_\perp\gamma_5} \widetilde H(u,\xi)
-\frac{\Delta_\perp^\mu}{2\xi} \biggl[u\frac{\partial}{\partial
u}+\xi\frac{\partial}{\partial \xi}\biggl]\widetilde F(u,\xi) \, .
\ee
\be{Wpm} W_{\pm}(x,u,\xi)&=& \frac12\biggl\{
\theta(x>\xi)\theta(u>x)-\theta(x<\xi)\theta(u<x)
 \biggl\}\frac1{u-\xi} \nonumber
\\[4mm]&&\mskip-10mu \pm\frac12\biggl\{
\theta(x>-\xi)\theta(u>x)-\theta(x<-\xi)\theta(u<x)
\biggl\}\frac1{u+\xi} . \ee
The flavor dependence in the amplitude can be easily restored by a
substitution~:
\be{flav}
F_\mu \, (\widetilde F_\mu) \to
\sum_{q=u,d,s, \dots}e_q^2\ F_\mu^q \, (\widetilde F_\mu^q) \, .
\ee

The light-cone expansion of the matrix elements (\ref{Voper},\ref{Aoper})
can be derived in a similar manner. To the
twist-3 accuracy needed for the present  calculation the result is
\be{VLCexp}
\langle P+\Delta/2|
\pbar(x)\gamma^\sigma\p (y)
 |P-\Delta/2\rangle &=& \int_{-1}^1 du e^{i(Px)(u-\xi)-i(Py)(u+\xi)}
\left\{
P^\sigma F(u,\xi)+F_\perp^\sigma(u,\xi)+\right.
\nonumber \\ &&\left.
\hspace*{-1cm} \frac12 iP^\sigma (x+y)_\rho\Delta_\perp^\rho F(u,\xi)+
iP^\sigma(x-y)_\rho G_1^\rho(u,\xi)
\right\}
\ee
\be{ALCexp}
 \langle P+\Delta/2|
\pbar(x)\gamma^\sigma\gamma_5\p (y)
 |P-\Delta/2\rangle &=&
\int_{-1}^1 du e^{i(Px)(u-\xi)-i(Py)(u+\xi)}
\left\{
P^\sigma \widetilde F(u,\xi)+\widetilde F_\perp^\sigma(u,\xi)+\right.
\nonumber \\ &&\left.
\hspace*{-1cm} \frac12 iP^\sigma (x+y)_\rho\Delta_\perp^\rho\widetilde
F(u,\xi)+
iP^\sigma(x-y)_\rho\tilde G_1^\rho(u,\xi)\, ,
\right\}
\ee
where now both $x$ and $y$ are off the light-cone. New
distributions
$G_1^\rho(u,\xi)$ and $\tilde G_1^\rho(u,\xi)$ can be expressed
through the skewed distributions \re{def:FWW}, \re{def:tFWW}~:
\be{G1}
G_1^\sigma (u,\xi)&=& u F_\perp^\sigma(u,\xi)-
\xi\, i\eps_\perp^{\sigma k}\widetilde F_{\perp k}(u,\xi)
-\frac12 i\eps_\perp^{\sigma k}\Delta_{\perp k}\widetilde F(u,\xi)
\ee
\be{tG1}
\tilde G_1^\sigma (u,\xi)&=& u\widetilde F_\perp^\sigma(u,\xi)-
\xi\, i\eps_\perp^{\sigma k} F_{\perp k}(u,\xi)
-\frac12 i\eps_\perp^{\sigma k}\Delta_{\perp k} F(u,\xi)
\ee
Keeping in mind calculation of the L$\rightarrow$T amplitude it is
useful to recall symmetry properties of skewed parton
distributions. Symmetry properties of twist-2 distributions
$H(u,\xi),\, E(u,\xi)$, $\widetilde H(u,\xi),\,\widetilde
E(u,\xi)$ with respect to interchange $u\leftrightarrow -u$ are
given by:
\be{Hsym}
H(-u,\xi)=-H(u,\xi),\, E(-u,\xi)=-E(u,\xi),\,
\\
\widetilde H(-u,\xi)= \widetilde H(u,\xi),\,
\widetilde E(-u,\xi)=\widetilde E(u,\xi).
\ee
As it follows,
\be{asym}
F(-u,\xi)=- F(u,\xi),\, F_{\perp \mu}(-u,\xi)=-F_{\perp \mu}(u,\xi)
\\
\widetilde F(-u,\xi)=\widetilde F(u,\xi), \,
\widetilde F_\perp^\mu(-u,\xi)=\widetilde F_\perp^\mu(u,\xi)
\ee
In the following we shall often need Fourier transforms of the matrix
elements \re{VLCexp} and \re{ALCexp}. For example, for the vector
matrix element one finds:
\be{VM}
&&\int \frac {d^4k'}{(2\pi)^4}e^{-i (k'y)}\int \frac
{d^4k}{(2\pi)^4}e^{-i (kx)}
\langle P+\Delta/2|
\pbar(y+\frac12 x)\gamma^\sigma\p (y-\frac12 x)
 |P-\Delta/2\rangle = \nonumber
\\ && \hspace{-2cm}
\int_{-1}^1 du
\left\{
P^\sigma F(u,\xi)+F_\perp^\sigma(u,\xi)
-P^\sigma \left(F(u,\xi) \Delta_\perp^\rho \frac{\partial}{\partial k'_\rho }
+ G_1^\rho(u,\xi)\frac{\partial}{\partial k_\rho }\right)
\right\}
\delta^{(4)}(2 \xi P + k')\, \delta^{(4)} (uP-k)\nonumber \\
\ee
Similar result holds for the Fourier transform of the axial matrix element.

\section*{\normalsize \bf Covariant calculation of the $L\rightarrow
T$ amplitude in the WW approximation}

The aim of this section is to present a method of calculation
of the $L\rightarrow
T$ DVCS amplitude in the WW approximation which can be efficiently
applied at the NLO.

The crucial simplification arises due to the fact that
according to \re{T} in the present
kinematics the  Lorentz indices $\mu$ and $\nu$ of the $L\rightarrow T$
amplitude point in the longitudinal, respectively transverse
directions. As it follows, $\mu$ can be
carried only by longitudinal
vectors $P$ and $q$ and $\nu$ has to be carried by transverse
components of other vectors present in the problem. Symbolically, one
can classify possible contributions as:
\be{tw3tensor}
(q \mbox{ or } P)^\mu\times \biggl\{ \Delta_\perp^\nu\, ,
i\epsilon^{\perp \nu}_\alpha\Delta_\perp^\alpha\, ,
F_\perp^\nu\, ,\,   G_1^\nu(u,\xi)\, ,\,\widetilde F_\perp^\nu,\,
\widetilde G_1^\nu(u,\xi)\,
  \biggr\}
\ee Hence, it is necessary to compute only those terms which can
be casted in the above form.

The starting point for calculation is the standard definition of
DVCS amplitude \re{T:def} which can be rewritten as:
\be{start} T^{\mu\nu}&=&i\int d^4x\ e^{-i (q+q')x/2}\langle
p'|T\left[J_{\rm e.m.}^\mu (x/2) J_{\rm
e.m.}^\nu(-x/2)\right]|p\rangle= \nonumber \\ && i\int d^4x\ e^{-i
(q+q')x/2}\int d^4y d^4z \langle
p'|\bar\psi(y)H(x,y,z)\psi(z)|p\rangle \ee
Function $H(x,y,z)$ is given by connected Feynman diagrams with
amputated external fermion legs, see Fig.~1.
\begin{figure}[t]
\unitlength1mm
\begin{center}
\hspace{0cm}  \inclfig{6}{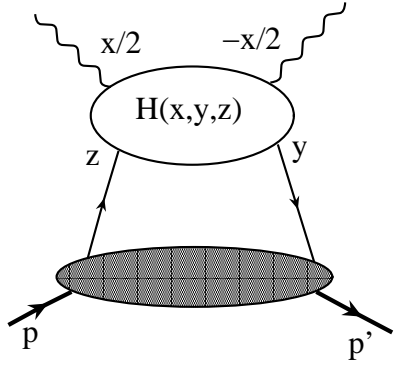}
\end{center}
\vspace{-0.5cm} \caption[dummy]{\small
 Graphical representation of the function $H(x,y,z)$
}
\end{figure}
Note that equation
\re{start} is valid only in the WW approximation i.e. if twist-3
contributions involving three points quark-gluon-quark matrix
elements are neglected. Using Fiertz identity we can further
transform \re{start} into:
\be{cont1}
T^{\mu\nu}= i\int d^4x\ e^{-i (q+q')x/2}\int d^4y \int d^4z
&&\hspace{-5mm}\biggl\{
 \tilde H_V^\sigma(x,y,z)
\langle p'|\bar\psi(y)\gamma_\sigma\psi(z)|p\rangle
\nonumber \\ &&
-
\tilde H_A^\sigma(x,y,z)
\langle p'|\bar\psi(y)\gamma_\sigma\gamma_5\psi(z)|p\rangle
\biggr\}\, .\nonumber \\
\ee
Here we have introduced notation
\be{H:def}
\tilde H_{V(A)}^\sigma(x,y,z)=\frac14{\rm tr}
\biggl\{
\gamma_\sigma(\gamma_\sigma\gamma_5) H(x,y,z)
\biggr\}
\ee
Using translational invariance and equations \re{VLCexp},\re{ALCexp} and
\re{VM} one easily obtains corresponding expression
in momentum space. The final result for the twist-3 part of
$T^{\mu\nu}$ which gives rise to the $L\rightarrow
T$ amplitude reads:
\be{finalT}
T^{\mu\nu}&=&\int_{-1}^{1}d u \int d^4 k \, \delta^4(uP-k)\int d^4 k'\,
\delta^4(2 \xi P + k')\int d^4 k^{''} \delta^4(q + \xi P -k^{''})
\nonumber \\ &&
\biggl\{ \left[ F_\perp^\sigma(u,\xi)+
P^\sigma \left( G_1^\rho(u,\xi)\frac{\partial}{\partial k^\rho } +
F(u,\xi) \Delta_\perp^\rho\Big(\frac{\partial}{\partial k^{'\rho} } -
\frac12 \frac{\partial}{\partial k^{''\rho} }\Big)\right)
\right]H_V^\sigma(k,k',k'')
\nonumber \\ &&
-\left[ \widetilde F_\perp^\sigma(u,\xi)+
P^\sigma\left( \tilde  G_1^\rho(u,\xi)\frac{\partial}{\partial k^\rho}
+\widetilde F(u,\xi) \Delta_\perp^\rho \Big(
\frac{\partial}{\partial k^{'\rho}} -
\frac12 \frac{\partial}{\partial k^{''\rho}}\Big)
 \right)
\right]H_A^\sigma(k,k',k'')
\biggr\}\, .\nonumber \\
\ee
Functions $H_{V(A)}^\sigma(k,k',k'')$ are Fourier transforms
of functions $\tilde H_{V(A)}^\sigma(x,y,z)$:
\be{H}
H_{V(A)}^\sigma(k,k',k'')=i\int d^4x\ e^{-i k'' x}\int d^4y e^{iy(k+k'/2)}
\int d^4z e^{-iz(k-k'/2)}\tilde H_{V(A)}^\sigma(x,y,z)\, .
\ee
Resulting flow of momenta is shown in Fig.~2.
\begin{figure}[t]
\unitlength1mm
\begin{center}
\hspace{0cm}  \inclfig{6}{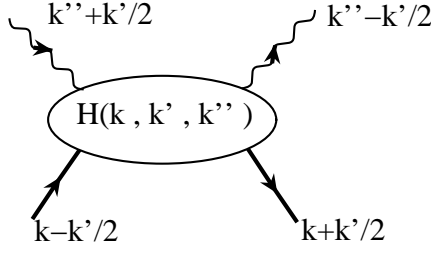}
\end{center}
\vspace{-0.5cm} \caption[dummy]{\small
 Momenta flow for Feynman diagrams of $H(k,k',k'')$
}
\end{figure}
Due to $\delta$-functions in \re{finalT}
one can put $k=uP$, $k' = -2\xi P$ and $k'' = q + \xi P$ after
differentiation. Thus, calculation of the
twist-3 DVCS amplitude in the WW approximation has been reduced to the
calculation of hard parton diagrams and their derivatives with respect
to external momenta.

Above considerations are quite general and valid at any order of perturbation
theory if the factorization is not broken.
Let us now illustrate them with the example of calculation of the LO
contribution to the twist-3
$L\rightarrow
T$ amplitude \re{T}. In this case the hard amplitudes
$H_{V(A)}^\sigma(k,k',k'')$ are given by
diagrams shown in Fig.~3.
\begin{figure}[t]
\unitlength1mm
\begin{center}
\hspace{0cm}  \inclfig{12}{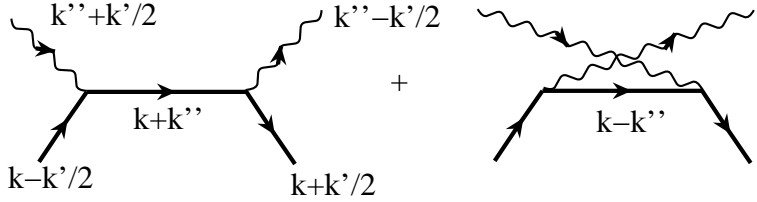}
\end{center}
\vspace{-0.5cm} \caption[dummy]{\small
 Leading order contribution to $H(k,k',k'')$
}
\end{figure}
Simple calculation gives:
\be{HBorn} H_{V(A)}^\sigma(k,k',k'')=\frac14 \frac{{\rm
tr}\biggl\{\gamma_\sigma(\gamma_\sigma\gamma_5)
\gamma^\mu(k+k'')\gamma^\nu\biggr\} } {(k+k'')^2+i0} \, , \ee
i.e. in the Born approximation $H_{V(A)}^\sigma(k,k',k'')$ does not
depend on $k'$. Note that one need to evaluate only one diagram as
contribution arising from the crossed one can be obtained
using symmetry considerations and \re{asym}.
Explicit calculation according to \re{finalT} yields:
\be{TLO1}
T^{\mu\nu}_{0+}=
\frac{1}{2(Pq)}\int_{-1}^1 du
\frac{1}{u+\xi-i0}\biggl\{
\frac12 \Delta_\perp^\nu P^\mu F(u,\xi) +
[(u+\xi)P^\mu-q'^\mu]F_\perp^\nu(u,\xi)
+
P^\mu G_1^\nu(u,\xi)
\nonumber \\
+\frac12 i\eps_{\perp k}^{\nu} \Delta_\perp^k P^\mu\widetilde F(u,\xi) -
[(u+\xi)P^\mu+q'^\mu]i\eps_{\perp k}^\nu\widetilde F_\perp^k (u,\xi)
+
P^\mu i\eps_{\perp k}^\nu \tilde G_1^k (u,\xi)
\biggr\}\, .\nonumber \\
\ee
Using the identity which follows from  \re{G1}, \re{tG1}:
\be{G1sum}
G_1^\nu(u,\xi)+\frac12 \Delta_\perp^\nu F(u,\xi)
+i\eps_{\perp k}^\nu \tilde G_1^k (u,\xi)-
\frac12 i\eps_{\perp k}^{\nu} \Delta_\perp^k \widetilde F(u,\xi)
= (u-\xi)\biggl[
F_\perp^\nu(u,\xi)+i\eps_{\perp k}^\nu\widetilde F_\perp^k (u,\xi)
\biggl]\nonumber \\
\ee
one can easily cast \re{TLO1} into the form
\be{TLO2}
T^{\mu\nu}_{0+}=
 -\frac{1}{2(Pq)}(4\xi P^\mu+q^\mu )\int_{-1}^1 du
\frac{1}{u+\xi-i0}
\biggl\{
F_\perp^\nu(u,\xi)+i\eps_{\perp k}^\nu\widetilde F_\perp^k (u,\xi)
\biggr\}+
\frac{1}{2(Pq)}P^\mu \int_{-1}^1 du F_\perp^\nu(u,\xi)\,.\nonumber \\
\ee
Taking into account the contribution from the crossed diagram one
finds that the term proportional to the integral
$\int_{-1}^1 du F_\perp^\nu(u,\xi)$ cancels out and the final
expression reads:
\be{Tres}
T^{\mu\nu}_{0+}&=&- \frac{1}{2(Pq)}(4\xi P^\mu+q^\mu )\int_{-1}^1 du
\biggl\{
\frac{1}{u+\xi-i0}
\biggl[
F_\perp^\nu(u,\xi)+i\eps_{\perp k}^\nu\widetilde F_\perp^k (u,\xi)
\biggr]
-
\nonumber \\ &&
\frac{1}{\xi-u-i0}
\biggl[
F_\perp^\nu(u,\xi)-i\eps_{\perp k}^\nu\widetilde F_\perp^k (u,\xi)
\biggr]
\biggr\} \, ,
\ee
which is precisely  the expression given in \re{T}.

One comment is in order here. In the present calculation of the Wilson
coefficient one does not need to apply QCD equations of motion (EOM) like
it was done in Refs. \ci{Penttinen} and \ci{Anikin}.
EOM had been used at the previous stage to derive
the light-cone expansion of the matrix elements \re{VLCexp} and
\re{ALCexp}. After it is done, the appropriate convolution of matrix
elements with the hard amplitude according to \re{finalT} leads
directly to the twist-3 $L\rightarrow T$ Compton amplitude.

\section*{\normalsize \bf NLO corrections to the WW twist-3 amplitude }

Technique described in the previous section can be applied
without any modifications to the calculation of the NLO flavor
non-singlet, quark contribution to the $L\rightarrow T$ Compton
amplitude. The corresponding Feynman diagrams shown in Fig.~4 are the
same as for the NLO contributions to twist-2 amplitude.
Nevertheless, besides trivial technical complications due to
derivatives of corresponding diagrams w.r.t. external momenta, some
new subtleties appear here which deserve a detailed
discussion before final presentation of NLO results.
\begin{figure}[t]
\unitlength1mm
\begin{center}
\hspace{0cm}  \inclfig{16}{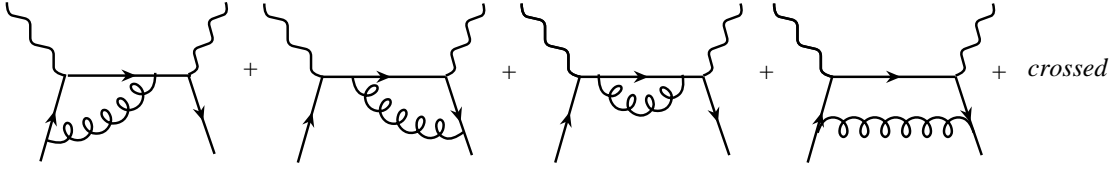}
\end{center}
\vspace{-0.5cm} \caption[dummy]{\small
  Feynmann diagrams for the NLO correction in WW approximation for $H(k,k',k'')$
}
\end{figure}

A convenient method for regularizing loop integrals arising at the
NLO is provided by dimensional regularization. We kept space-time
dimension $d = 4 - 2\eps$ and applied the $\overline{\rm MS}$
subtraction scheme. For the $\gamma_5$ matrix in $D$-dimension we
used the t'Hooft-Veltman definition in terms of four antisymmetric
gamma-matrices \footnote{Self-consistency of such requires to
perform additional finite renormalization of the local axial
current operator \ci{Larin}.}. Powerful check of the resulting
calculation is provided by the electromagnetic gauge-invariance of
the final answer. We computed all possible contributions according
to \re{tw3tensor} and check that in the sum of all diagrams only
such combination of momenta appears, $q^\mu+4\xi P^\mu$, which is
consistent with electromagnetic gauge-invariance.

So far, it is a standard discussion familiar from calculations of
higher-order coefficients in twist-2 amplitudes. New and
interesting element appears when one considers  a possible
structure of the NLO answer. Obviously, proportionality of the $L
\rightarrow T$ amplitude in \re{T} to $q^\mu + 4\xi P^\mu $  is
fixed by gauge-invariance arguments and does not change when one
goes from LO to NLO accuracy. On the other hand, the the LO
amplitude depends only on combinations of $F_\perp^\nu$ and
$\widetilde F_\perp^k$. Contributions involving $F$ and
$\widetilde F$,  which might in principle be present according to
\re{finalT}, cancel. This cancellation is, however, specific to
the Born approximation and does not hold at higher orders. We
therefore expect that the NLO expression for the $L \rightarrow T$
amplitude has the form:
\be{def:TNLO}
\hspace{-1cm}T^{ \mu \nu}_{0+}&=&
\frac{(q+4\xi P)^\mu}{2(Pq)} \left[g^{\nu
\alpha}_\perp+\frac{P^\nu\Delta_\perp^\alpha}{(Pq)}  \right]
\int_{-1}^1 dx\,\biggl\{
C_\perp(x,\xi;\mu^2/Q^2)
\biggl( F_{\perp\alpha}(x,\xi)+
i\epsilon^\perp_{\alpha \rho}\widetilde F^\rho_\perp
(x,\xi)\biggr) \nonumber
\\ &&
+ \frac{\alpha_s(\mu^2)}{2\pi}C_F
\left[\Delta_\perp^\alpha
C(x,\xi;\mu^2/Q^2)F(x,\xi)+i\epsilon^\perp_{\alpha \rho}
\Delta_\perp^\rho \widetilde C(x,\xi;\mu^2/Q^2) \widetilde F(x,\xi)
\right]
\biggr\}  \, ,\nonumber \\
&&C_\perp(x,\xi;\mu^2/Q^2) =
\frac{-2}{x+\xi}+\frac{\alpha_s(\mu^2)}{2\pi}C_F
C_\perp^{(1)}(x,\xi;\mu^2/Q^2) \, . \ee
Here we used symmetry properties of skewed parton distributions
\re{asym} in order to rewrite the contribution from crossed diagrams
in terms of direct ones.

Note now that term containing $\Delta_\perp^\alpha F(x,\xi)$ may
interfere with a corresponding piece of the $L \rightarrow L$
amplitude $T^{ \mu \nu}_{00}$ which appears at the NLO and
describes the twist-2 transition between longitudinally polarized
initial and final photons. In the massless QCD the corresponding
contribution to the hadronic tensor $T^{ \mu \nu}_{00}$ has
logarithmic singularity $\sim \ln \left[q^2/q'^2 \right]$ as the
virtuality of the final photon tends to zero. Of course, the $L
\rightarrow L$ scattering amplitude obtained by contraction of
$T^{ \mu \nu}_{00}$ with polarization vectors of initial and final
photons vanishes in this limit. In the Feynman gauge the
contribution to $T^{ \mu \nu}_{00}$ arises from the box diagram
\ci{MPS}. From the general parametrization, valid for non-zero
$q'^2$:
\be{TLL}
T^{ \mu \nu}_{00}=\frac1{(Pq)^2}\left(q^\mu(Pq)-P^\mu q^2 \right)
\left(q'^\nu(Pq)-P^\nu q'^2 \right){\rm  A}_{LL}
\ee
one finds that formally  $T^{ \mu \nu}_{00}$ has a twist-3 part which
can indeed contribute to $T^{\mu\nu}$ with index $\mu$ transverse and
$\nu$ longitudinal. Retaining only the contribution singular as
$q'^{2}$ goes to zero, one finds:
\be{extra}
T^{ \mu \nu}_{00}|_{tw-3}&=&(q+4\xi P)^\mu (-\Delta_\perp^\nu)
{\rm  A}_{LL}\, , \nonumber \\
{\rm A}_{LL}&=&\frac1{2(Pq)}\frac{\alpha_s(\mu^2)}{4\pi}C_F
\int_{-1}^1 dx F(x,\xi) \biggl[\frac{\ln \left[q^2/q'^2
\right]}{(\xi+x)2\xi}  + \frac{\ln \left[\frac{\xi+x}{2\xi}
\right]}{(\xi+x)(\xi-x)} \biggr]\, . \ee
Note that expression for ${\rm A}_{LL}$ is finite as the space-time
dimension $d=4$ but instead it
has singularity when $q'^2\to 0$. We regulated it by a small non-zero
value of $q'^2$. In order to obtain the NLO $L \rightarrow T$
amplitude the contribution \re{extra} must be subtracted from result of
calculation of diagrams in Fig.~4 in $d=4 - 2\eps$  dimensions and
with $q'^2 \neq 0$. In practice, in the Feynman gauge only the box
graph gives contribution singular as $q'^2 \to 0$.

Explicit calculation according to the discussion above gives in $d=4
- 2 \eps$ dimensions the
following
result for the coefficient functions $C_\perp^{(1)}(x,\xi;\mu^2/Q^2)$,
$C(x,\xi;\mu^2/Q^2)$ and
$\widetilde C(x,\xi;\mu^2/Q^2)$ in \re{def:TNLO}:
\be{Cperp}
C_\perp^{(1)}(x,\xi;\mu^2/Q^2)&=& \frac{1}{x+\xi}\biggl\{ -\frac1\eps \biggl[ 1+
2\ln \left[\frac{\xi+x}{2\xi} \right]\biggr](1+\eps\ln[\mu^2/Q^2])
\nonumber \\ &&
-5-2\ln \left[\frac{\xi+x}{2\xi} \right]+
4\ln \left[\frac{\xi-x}{2\xi} \right]-
\frac{\xi-x}{\xi+x}\ln \left[\frac{\xi-x}{2\xi} \right]+
\ln \left[\frac{\xi+x}{2\xi} \right]^2 \biggr\}
\, , \nonumber \\
\ee
\be{C}
C(x,\xi;\mu^2/Q^2)&=&\frac1\eps\frac1{\xi(\xi+x)}(1+\eps\ln[\mu^2/Q^2])+
\frac1{\xi^2-x^2} \biggl( 2\frac x\xi -\ln
\left[\frac{\xi+x}{2\xi} \right]+ \frac{2x}{\xi-x}\ln
\left[\frac{\xi+x}{2\xi} \right]
\biggr)\, , \nonumber \\
\ee
\be{tC} \widetilde
C(x,\xi;\mu^2/Q^2)&=&\frac1\eps\frac1{\xi(\xi+x)}(1+\eps\ln[\mu^2/Q^2])+
\frac1{\xi^2-x^2} \biggl( 2-\ln \left[\frac{\xi+x}{2\xi} \right]-
\frac{2\xi}{\xi-x}\ln \left[\frac{\xi+x}{2\xi} \right]
\biggr) \, . \nonumber \\
\ee

Two important comments are of order here. First, in
the above equations it is understood that $\xi$ has
infinitesimally small imaginary part $\xi \to \xi-i\eps$.
Moreover, as the combination $F_{\perp\alpha}+
i\epsilon^\perp_{\alpha \rho}\widetilde F^\rho_\perp$ is continuous at
the point $x=-\xi$, the resulting integrals in \re{def:TNLO} are well
defined, in complete analogy with the situation in the LO
amplitude. However, in order to complete proof of factorization of the $L
\rightarrow T$ amplitude at
the NLO one has to demonstrate more, namely that the structure of
singular terms is
indeed such that they can be absorbed, following the standard
procedure, into renormalization of the LO
$L \rightarrow T$ amplitude.

In order to show this one has to derive one-loop evolution
equations for the combination
$F_{\perp\alpha}+i\epsilon^\perp_{\alpha \rho}\widetilde
F^\rho_\perp$ which appears at the LO in \re{Tshort}. One
possibility is to use definitions \re{def:Fperp},\re{def:tFperp}
and the well-known evolution equations of the twist-2 SPD's $F$
and $\widetilde F$ see, for instance,  \ci{Rad97}. Another way
is to compute directly the one-loop evolution of the transverse
components of the light-cone matrix elements \re{Fdef} and
\re{Ftlddef}. The result is:
\be{evol}
\mu^2\frac{d}{d\mu^2}
( F_{\perp\alpha}+i\epsilon^\perp_{\alpha \rho}\widetilde F^\rho_\perp)
(x,\xi) &=&
\frac{\alpha_s(\mu^2)}{2\pi}C_F
\int_{-1}^1 d u\biggl[
 V_+(u,x,\xi)
( F_{\perp\alpha}+i\epsilon^\perp_{\alpha \rho}\widetilde
F^\rho_\perp)(u,\xi)
\nonumber \\ &&
+U_+(u,x,\xi)(\Delta_\perp^\alpha F+i\epsilon^\perp_{\alpha \rho}
\Delta_\perp^\rho  \widetilde F)(u,\xi)
\biggr]\, .
\ee
The evolution kernels are defined as:
\be{Vplus}
V_+(u,x,\xi)&=& \biggl[\theta(\xi<x<u)-\theta(u<x<\xi) \biggr]
\left(\frac{1}{u-x}\right)_+
\\ && +
\biggl[\theta(-\xi<x<u)-\theta(u<x<-\xi)\biggr]
\left(
\frac{1}{u-x}
\frac{(\xi+x)^2}{(\xi+u)^2}
\right)_+\, ,
\ee
\be{Uplus}
U_+(u,x,\xi)&=& \biggl[\theta(\xi<x<u)-\theta(u<x<\xi) \biggr]
\left(\frac{(\xi+x)}{4\xi^2(u-\xi)}\right)_+
\\ && -
\biggl[\theta(-\xi<x<u)-\theta(u<x<-\xi)\biggr]
\left(\frac{(\xi+x)(3\xi+u)}{4\xi^2(u+\xi)^2}\right)_+ \, .
\ee
Here the plus prescription is to be understood as
\be{pluspr}
\left(X(u,x,\xi)\right)_+=X(u,x,\xi)-\delta(x-u)\int dz X(u,z,\xi)
\ee
As one can see the  evolution equation of the WW contribution is not
homogenous, i.e. the RHS of \re{evol} receives contributions which are
not given in terms of the combination of GPD's on the LHS. Precisely
these terms generate contributions to the one-loop renormalization of
the LO amplitude which exactly match singular terms arising from
``bare'' coefficients $C$ and $\widetilde C$.
Indeed, with the help of equations \re{Vplus} and \re{Uplus} one finds
that convolution integrals
of the tree level coefficient function with kernels $V_+$ and $U_+$
do reproduce the pole contributions in \re{Cperp}, \re{C} and \re{tC}
as it should be:
\be{poles} \frac{\alpha_s(\mu^2)}{4\pi}C_F\int_{-1}^1 d
x\frac{-2}{(\xi+x)}V_+(u,x,\xi)&=&
-\frac{\alpha_s(\mu^2)}{2\pi}C_F \frac1{(\xi+u)}\left( 1+
2\ln\left[\frac{\xi+u}{2\xi} \right]\right)\, ,
\\
\frac{\alpha_s(\mu^2)}{2\pi}C_F\int_{-1}^1 d
x\frac{-2}{(\xi+x)}U_+(u,x,\xi)&=&
\frac{\alpha_s(\mu^2)}{2\pi}C_F\frac1\xi\frac1{(\xi+u)} \, .
\ee

Now, combining equations \re{Cperp}, \re{C}, \re{tC} with \re{poles}
one obtains the NLO Wilson coefficients in their final form:
\be{Cperpf}
C_\perp^{(1)}(x,\xi;\mu^2/Q^2)&=& \frac{1}{x+\xi}\biggl\{ - \biggl( 1+
2\ln \left[\frac{\xi+x}{2\xi} \right]\biggr) \ln[\mu^2/Q^2]
\nonumber \\ &&
-5-2\ln \left[\frac{\xi+x}{2\xi} \right]+
4\ln \left[\frac{\xi-x}{2\xi} \right]-
\frac{\xi-x}{\xi+x}\ln \left[\frac{\xi-x}{2\xi} \right]+
\ln \left[\frac{\xi+x}{2\xi} \right]^2 \biggr\}
\, , \nonumber \\
\ee
\be{Cf}
C(x,\xi;\mu^2/Q^2)&=& \frac1{\xi(\xi+x)}\ln[\mu^2/Q^2]+
\frac1{\xi^2-x^2} \biggl( 2\frac x\xi -\ln
\left[\frac{\xi+x}{2\xi} \right]+ \frac{2x}{\xi-x}\ln
\left[\frac{\xi+x}{2\xi} \right]
\biggr)\, , \nonumber \\
\ee
\be{tCf} \widetilde
C(x,\xi;\mu^2/Q^2)&=& \frac1{\xi(\xi+x)}\ln[\mu^2/Q^2]+
\frac1{\xi^2-x^2} \biggl( 2-\ln \left[\frac{\xi+x}{2\xi} \right]-
\frac{2\xi}{\xi-x}\ln \left[\frac{\xi+x}{2\xi} \right]
\biggr) \, . \nonumber \\
\ee

The above equations are our main result. They explicitly show that
the NLO correction to the $L \rightarrow T$ amplitude can be put
into a factorizable form in a self-consistent way.

\section*{\normalsize \bf Conclusions }

In this paper we have demonstrated factorizability of the NLO
correction to the twist-3, Wandzura-Wilczek, flavor non-singlet
DVCS amplitude corresponding to scattering of longitudinal photon
off nucleon. We have shown that, similarly to the LO, the
singularity structure of the corresponding Wilson coefficients is
such that the convolution integrals with SPD's are well defined.
After the interference with the twist-2 LL amplitude is properly
taken into account all IR singular terms arising in the
calculation of the $L \rightarrow T$ amplitude can be absorbed
into renormalization of the LO result in a usual manner. Besides
the phenomenological applications of new Wilson coefficients found
in this paper our result strongly suggests that factorization
holds for the twist-3, $L \rightarrow T$ DVCS amplitude beyond LO
and possibly to all orders in the $\alpha_S$ expansion.

\section*{\normalsize \bf Acknowledgments }
LM wants to acknowledge kind hospitality extended to him during
numerous visits to Regensburg and Bochum Universities. Research
reported in this paper has been supported in part by Polish-German
PAN-DFG grant, Kovalevskaja Program of
Alexander von Humboldt Foundation, by the BMBF and by the
COSY-Julich project.

\end{document}